\documentclass[preprint,3p,sort&compress,times]{elsarticle}

\usepackage[T1]{fontenc}
\usepackage{graphicx}
\usepackage{latexsym}
\usepackage{amsmath, amsthm, amsfonts, amssymb}
\usepackage{stackrel}
\usepackage{epstopdf,comment}

\journal{Journal of \LaTeX\ Templates}

\begin{document}

\begin{frontmatter}

\title{Hermite--Gaussian model for quantum states}
\author[uba]{Marcelo Losada\corref{cor1}}
\ead{marcelolosada@yahoo.com}
\author[ufba,inct]{Ignacio S. Gomez}
\ead{nachosky@fisica.unlp.edu.ar}
\author[iflp]{Federico Holik}

\cortext[cor1]{Corresponding author}
\address[uba]{Universidad de Buenos Aires - CONICET Ciudad Universitaria, 1428 Buenos Aires, Argentina.}
\address[ufba]{Instituto de F\'{i}sica, Universidade Federal da Bahia,
         Rua Barao de Jeremoabo, 40170-115 Salvador--BA, Brazil}
\address[inct]{National Institute of Science and Technology for Complex Systems,
        Rua Xavier Sigaud 150, Rio de Janeiro 22290-180,
        Brazil}
\address[iflp]{IFLP, UNLP, CONICET, Facultad de Ciencias Exactas,
               Calle 115 y 49, 1900 La Plata, Argentina}


\begin{abstract}
In order to characterize quantum states within the context of information geometry, we propose
a generalization of the Gaussian model, which we called the \emph{Hermite--Gaussian model}. We obtain the Fisher--Rao metric and the scalar curvature for this model, and we show its relation with the one-dimensional quantum harmonic oscillator.
Moreover, using this model we characterize some families of states of the quantum harmonic oscillator. We find that for eigenstates of the Hamiltonian, mixtures of eigenstates and even or odd superpositions of eigenstates the associated Fisher--Rao metrics are diagonal.
\end{abstract}


\end{frontmatter}

\section{Introduction}




The information geometry approach \cite{fisher,frieden,jaynes,rao, rao2, weinhold,
ruppeiner, amari, ingarden, janyszek} studies the differential
geometric structure of statistical models. A statistical model
consists of a family of probability distribution functions (PDFs)
parameterized by continuous variables. In order to endow these
models with a geometric structure, it is necessary to define the
Fisher--Rao metric \cite{rao}, which in turn, is linked with the
concepts of entropy and Fisher information. Once we have a
statistical manifold, the main goal of the information geometry
approach is to characterize the family of PDFs using geometric
quantities, like the geodesic equations, the Riemann curvature
tensor, the Ricci tensor or the scalar curvature.

The geometrization of thermodynamics and statistical mechanics are
some of the most important achievements in this field, expressed mainly by the
foundational works of
Gibbs \cite{gibbs}, Hermann \cite{hermann},
Weinhold \cite{weinhold}, Mruga{\l}a \cite{mrugala},
Ruppeiner \cite{ruppeiner2}, and Carathe\'{o}dory \cite{carateodory}.
These investigations lead to the Weinhold and Ruppeiner
geometries, where a
a Riemannan metric tensor in the space of
thermodynamic parameters is provided and a notion of distance
between macroscopic states is obtained.
However, the
utility of information geometry is not only limited to those areas.
For instance, it has been applied in quantum mechanics leading to a
quantum generalization of the Fisher--Rao metric \cite{bures}, and
recently, also in nuclear plasmas \cite{geert,geert2}. Moreover,
generalized extensions of the information geometry approach to the
non-extensive formulation of statistical mechanics \cite{tsallis}
have been also considered \cite{abe,naudts,portesi,portesi2}. Applications of
information geometry to chaos can be also performed by considering
complexity on curved manifolds
\cite{cafaro,cafaro2,cafaro3,cafaro4,cafaro5}, leading to a criterion for
characterizing global chaos on statistical manifolds: the more
negative is the curvature, the more chaotic is the dynamics; from which
some consequences concerning dynamical systems have been explored
\cite{nacho1}.
More generally, the curvature has been proved to be
a quantifier which measures interactions in thermodynamical systems,
where the positive or negative sign corresponds to repulsive or
attractive correlations, respectively \cite{ruppeiner}.

Motivated by previous works
of some of us \cite{nacho1,nacho2},
we propose a generalization of the Gaussian model
which we call the \emph{Hermite--Gaussian model}, and we show its
relation with the one-dimensional quantum harmonic oscillator.
The application of information geometry techniques to the study of
quantum harmonic oscillators can be useful in many applications. For
example, the translational modes in a quantum ion trap are quantum
harmonic oscillators that need to be characterized and controlled in
order to avoid coherence losses. Our contribution may serve as a
tool for the characterization of unknown parameters in those
scenarios.
Furthermore, the present work can be also
considered as a continuation of the recent Cafaro's
program \cite{cafaro,cafaro2,cafaro3,cafaro4,cafaro5}
of global characterization of dynamics on
curved statistical manifolds generated by Gaussian models.

The paper is organized as follows. In Section II, we review the main
features of the information geometry approach. In Section III, we
present the Hermite--Gaussian model, we obtain the Fisher--Rao
metric and the scalar curvature for this model, and we show its
relation with the one-dimensional quantum harmonic oscillator.
Moreover, we use this model to characterize some families of states
of the quantum harmonic oscillator. We focus in three different
families of states: Hamiltonian eigenstates, mixtures of eigenstates
and superposition of eigenstates. Finally, in Section IV, we present
the conclusions and some future research directions.

\section{Information geometry}


The information geometry approach studies the differential geometric
structure possessed by families of probability distribution
functions (PDFs). In this section we introduce the general features
of this approach, which will be used in the next sections. The
presentation is based on the book of S. Amari and H. Nagaoka
\cite{amari}.

\subsection{Statistical models}\label{subsec statistical models}

Information geometry applies techniques of differential geometry to
study properties of families of probability distribution functions
parameterized by continuous variables. These families are called
\emph{statistical models}. More specifically, a statistical model is
defined as follows. We consider the probability distribution
functions defined on $X \subseteq \mathbb{R}^n$, i.e., the functions
$p:X\rightarrow\mathbb{R}$ which satisfy

\begin{eqnarray}\label{2-1}
p(x)\geq0, ~~~ \text{and } ~~~ \int_X p(x)dx =1.
\end{eqnarray}

When $X$ is a discrete set the integral must be replaced by a sum. A
statistical model is a family $S$ of probability distribution
function on $X$, whose elements can be parameterized by appealing to
a set of $m$ real variables, i.e.,

\begin{eqnarray}\label{2-2}
S=\left\{  p_{\theta}(x)=p(x|\theta) \ \middle| \ \theta=(\theta^1,\ldots,\theta^m)\in \Theta \subseteq \mathbb{R}^m  \right\},
\end{eqnarray}

\noindent with $\theta\mapsto p_{\theta}$ an injective mapping. The
dimension of the statistical model is given by the number of real
variables used to parameterized the family $S$.

When statistical models are applied to physical systems, the
interpretation of $X$ and $\Theta$ is the following. $X$ represents
the microscopic variables of the system, which are typically
difficult to determine, for instance the positions of the particles
of a gas. $\Theta$ represents the macroscopic variables of the
system, which can be easily measured. The set $X$ is called the
\emph{microspace} and the variables $x \in X$ are the
microvariables. The set $\Theta$ is called the \emph{macrospace} and
the variables $\theta^1,\ldots,\theta^m$ are the macrovariables.

Given a physical system, we can define many statistical models.
First, we have to choose the microvariables to be considered, and
then we have to choose the macrovariables which parametrized the
PDFs defined on the microspace. All statistical models are equally
valid, but no all of them are equally useful. In general, the choice
of the statistical model would be based in pragmatic considerations.

\subsection{Geometric structure of statistical models}\label{subsec statistical manifold}

In order to apply differential geometry to statistical models, it is necessary to endow them with a metric structure.
This is accomplished by means of the Fisher--Rao metric
\begin{eqnarray}\label{fisherao}
\bold{I}=I_{ij}=\int_X dx \ p(x|\theta)\frac{\partial \log
p(x|\theta)}{\partial \theta^i }\frac{\partial \log
p(x|\theta)}{\partial \theta^j },  \ \ \ \ \ \ \ \ \ \ \ \
i,j=1,\ldots,m.
\end{eqnarray}

\noindent The metric tensor $\bold{I}$ gives to the macrospace a
geometrical structure. Therefore, the family $S$ results to be a
statistical manifold, i.e., a differential manifold whose elements
are probability distribution functions.

From the Fisher--Rao metric, we can obtain the line element between two nearby PDFs with parameters $\theta^i+d\theta^i$ and $\theta^i$
\begin{eqnarray}\label{linelement}
ds=\sqrt {I_{ij}d\theta^id\theta^j},  \ \ \ \ \ \ \ \ \ \ \ \ i,j=1,\ldots,m. \nonumber
\end{eqnarray}

Using the metric tensor we can obtain the geodesic equations for the
macrovariables $\theta_i$ along with relevant geometrical
quantities, like the Riemann curvature tensor, the Ricci tensor or
the scalar curvature.

\begin{eqnarray}
\textit{Geodesic equations:} & \frac{d^2\theta_k}{d^2\tau}+\Gamma_{ij}^k\frac{d\theta_i}{d\tau}\frac{d\theta_j}{d\tau}=0,   \label{geodesic} \\
\textit{Christoffel symbols:} &  \Gamma_{ij}^k=\frac{1}{2}I^{im}\left(I_{mk,l}+I_{ml,k}-I_{kl,m} \right), \label{christoffel} \\
\textit{Riemman curvature tensor:} & R_{iklm}=\frac{1}{2}\left(I_{im,kl}+I_{kl,im}-I_{il,km}-I_{km,il}\right)+I_{np}\left(\Gamma_{kl}^n\Gamma_{im}^p-\Gamma_{km}^n\Gamma_{il}^p \right), \label{riemann} \\
\textit{Ricci tensor:} & R_{ik}=I^{lm}R_{limk}, \label{ricci} \\
\textit{Scalar curvature:} & R=I^{ik}R_{ik}. \label{scalar}
\end{eqnarray}

The comma in the sub-indexes denotes the partial derivative
operation (of first and second orders), $I^{kl}$ is the inverse of
$I_{ij}$, and $\tau$ is a parameter that characterizes the geodesic
curves.


Moreover, the Fisher--Rao metric gives information about the
estimators of the macrovariables. Given an unbiased estimator
$\textbf{T}= (T_1, ..., T_m)$ of the parameters $(\theta_1, ...,
\theta_m)$, i.e., $E\left(\textbf{T}\right)= (\theta_1, ...,
\theta_m)$, the Cram\'er--Rao bound gives a lower bound for the
covariance matrix of $\textbf{T}$,

\begin{equation}\label{cramer rao}
  \text{cov}\left( \bold{T}\right)\geq \bold{I}^{-1},
\end{equation}

\noindent where the matrix inequality $A\geq B$ means that the
matrix $A-B$ is positive semi-definite. In particular, this relation
gives bounds for the variance of the unbiased estimators $T_i$,

\begin{equation}\label{cramer rao}
  \text{var}\left( T_i\right)\geq   \{ \bold{I}^{-1}\}_{ii},
\end{equation}

\noindent This bound is important when looking for optimal
estimators. In what follows, we present an important statistical
model used in the geometry information approach, the Gaussian model.

\subsection{Gaussian model}\label{subsec gaussianmodel}

One of the most relevant statistical models used in the geometry
information approach is the \emph{Gaussian model}. The reason for
that is the wide versatility of this model for describing multiple
phenomena. The Gaussian model is obtained by choosing the family $S$
as the set of multivariate Gaussian distributions. For instance, if
$(x_1,\ldots,x_n)\in \mathbb{R}^n$ are the microvariables and there
is no correlations between them, then
$(\mu_1,\ldots,\mu_n,\sigma_1,\ldots,\sigma_n)\in
\mathbb{R}^n\times\mathbb{R}_+^n$ are the set of macrovariables,
where $\mu_i$ and $\sigma_i^2$ correspond to the mean value and the
variance of the microvariable $x_i$ .

If we consider only one microvariable $x$, the Gaussian model is
given by the following probability distribution functions

\begin{eqnarray}\label{gaussian model}
p(x|\mu,\sigma)=\frac{1}{\sqrt{2\pi}\sigma}e^{-\frac{(x-\mu)^2}{2\sigma^2}},
\end{eqnarray}

\noindent which are parameterized by the mean value $\mu$ and the
standard deviation $\sigma$. From equations \eqref{fisherao} to
\eqref{scalar}, one can obtain the Fisher--Rao metric and the scalar
curvature of this model,

\begin{equation}\label{fisherao-gaussian}
I_{\alpha \beta}=\left(
\begin{array}{cc}
\frac{1}{\sigma^2} & 0 \\
0 & \frac{2}{\sigma^2} \\
\end{array}
\right) ~~~~~ \text{with} ~~~~~ \alpha, \beta = \mu, \sigma,
\end{equation}
\begin{equation}\label{curvatura-gaussian}
R=-1.
\end{equation}

\noindent The Gaussian model is a curved manifold with constant
curvature. In some contexts, the negative value of the curvature is
interpreted as modeling attractive interactions, like in an ideal
gas \cite{ruppeiner}.

In the next section, we are going to introduce a generalization of
the Gaussian model, based on the eigenstates of the harmonic
oscillator Hamiltonian.

\section{Hermite--Gaussian model}

We propose a generalization of the Gaussian model, called the
\emph{Hermite-Gaussian model}, which is motivated by the quantum
harmonic oscillator. Given the microspace $X = \mathbb{R}$ and the
macrospace $\Theta=\{(\mu,\sigma)\}$, we define for each $n$ the
Hermite--Gaussian model as the family of probability distribution
functions given by

\begin{eqnarray}\label{gaussian hermitemodel}
p_n(x|\mu,\sigma) =\frac{1}{\sqrt{2 \pi}\sigma}\, e^{-\frac{(x-\mu)^2}{2\sigma^2}}a_n^2 H^2_n\left(\frac{x-\mu}{\sqrt{2}\sigma}\right), ~~~~~~~a_n=\frac{1}{\sqrt{2^n n!}}.
\end{eqnarray}

\noindent In particular, if $n=0$, the Gaussian model is recovered.
The Fisher--Rao metric of the Hermite--Gaussian model takes the form

\begin{equation}\label{metrica nHS}
I^{(n)}_{\alpha \beta}=\int_X  \ \frac{1}{p(x|\mu, \sigma)} \partial_{\alpha}p(x|\mu, \sigma) \partial_{\beta}p(x|\mu, \sigma) dx , ~~~~~~ ~~~~~~\alpha,\beta= \mu, \sigma.
\end{equation}

\noindent and its explicit formula is the following (see
\ref{apendice2})

\begin{eqnarray}\label{fisher-eigenstate}
I^{(n)}_{\alpha \beta}=\left(
\begin{array}{cc}
\frac{2n+1}{\sigma^2} & 0 \\
0 & \frac{2(n^2+n+1)}{\sigma^2} \\
\end{array}
\right).
\end{eqnarray}

\noindent Taking into account that the scalar curvature is given by

\begin{equation}\label{curvature-eigenstate}
R^{(n)}= -\frac{1}{n^2+n+1},
\end{equation}

\noindent we can express the Fisher--Rao metric in terms of
$R^{(n)}$

\begin{eqnarray}\label{fisher-eigenstate-curvature}
I^{(n)}_{\alpha \beta}=\left(
\begin{array}{cc}
\frac{2n+1}{\sigma^2} & 0 \\
0 & -\frac{2}{ \sigma^2 R^{(n)}} \\
\end{array}
\right).
\end{eqnarray}

\noindent From the Fisher--Rao metric, we can compute the
Cram\'er--Rao bound for unbiased estimators of the parameters $\mu$
and $\sigma$. The lower covariance matrix of any pair of unbiased
estimators $T_1, T_2$ of the parameters $\mu, \sigma$, is given by

\begin{equation}
  \text{cov}\left( T_1, T_2\right)\geq \left(
\begin{array}{cc}
\frac{\sigma^2}{2n+1} & 0 \\
0 & -\frac{\sigma^2 R^{(n)}}{2} \\
\end{array}
\right).
\end{equation}

\noindent For the covariance of the estimators we obtain

\begin{align}
   \text{var} \left( T_1\right)&\geq \frac{\sigma^2}{2n+1}, \\
  \text{var} \left( T_2\right)&\geq \frac{\sigma^2 }{2 (n^2+n+1)} = -\frac{\sigma^2 R^{(n)} }{2}.
\end{align}

\noindent In what follows, we show the connection between the
Hermite--Gaussian model and the quantum harmonic oscillator. We use
these model to characterize the PDFs given by quantum states of the
harmonic oscillator. We focus on Hamiltonian eigenstates, mixtures
of eigenstates and superposition of eigenstates.

\subsection{Hamiltonian Eigenstates}

The relation between the Hermite--Gaussian model and the quantum harmonic oscillator is straightforward. We start considering the Hamiltonian of the harmonic oscillator
\begin{eqnarray}\label{hamiltonian}
\hat{H}=\frac{\hat{p}^2}{2m}+\frac{1}{2}m\omega_0^2 (\hat{x}-x_0)^2,
\end{eqnarray}
where $m$ is the mass, $\omega_0$ is the frequency, $x_0$ is the equilibrium position of the oscillator, and $\hat{x}$ and $\hat{p}$ are the position and momentum operators.
Its eigenstates $|n\rangle$ satisfy the time-independent Schr\"{o}dinger equation, $\hat{H}|n\rangle = E_n |n\rangle$, with $E_n = \hbar \omega_0 (n + \frac{1}{2})$. Moreover, the eigenstates satisfy orthogonality and completeness relations
\begin{equation}
\begin{split}
\langle n|m\rangle=\delta_{nm}\hspace{1cm} &\textrm{(orthogonality)}\nonumber    \\
\sum_{n=0}^{\infty}|n\rangle\langle n|=\hat{I} \hspace{1cm}&\textrm{(completeness)} \nonumber
\end{split}
\end{equation}
where $\hat{I}$ is the identity operator.

The wave function of the eigenstate $|n\rangle$, in the coordinate representation, is given by
\begin{eqnarray}\label{wave function}
\varphi_n(x)= \langle x|n\rangle= \frac{1}{\sqrt{\sqrt{2\pi}\sigma}}e^{-\frac{(x-\mu)^2}{4\sigma^2}}a_n H_n\left(\frac{x-\mu}{\sqrt{2}\sigma}\right),
\end{eqnarray}
with $\mu = x_0$, $\sigma^2= \frac{\hbar}{2 m \omega_0}$, and $a_n=\frac{1}{\sqrt{2^n n!}}$. Then, the PDF of the position operator for the eigenstate $|n\rangle$ is $P_n(x)=|\varphi_n(x)|^2$.

Therefore, if we consider the eigenstate $|n\rangle$ of an harmonic oscillator with parameters $\mu$ and $\sigma$, the PDF of the position operator $P_n(x)$ is equal to the probability distribution function  $p_n(x|\mu,\sigma)$ of the Hermite--Gaussian model, given in equation (\ref{gaussian hermitemodel}).
Moreover, the Fisher--Rao metric and the scalar curvature associated with the probability distribution function $P_n(x)$ are given in equations \eqref{fisher-eigenstate} and \eqref{curvature-eigenstate}, respectively.

It is important to remark that the Fisher--Rao metric is diagonal, and the scalar curvature is always negative and decreases with the quantum number $n$,
tending to zero in the limit of high quantum numbers. Moreover, from the Cram\'er--Rao bound we obtain that the minimal variance of the estimators of the parameter $\mu$ grows with $\sigma ^2$ and decreases with the eigenstate number, and the minimal variance of estimators of the parameter $\sigma$ also grows with $\sigma ^2$ but decreases with the square of the eigenstate number. Equivalently, the minimal variance of the estimators of $\sigma$ is proportional to the scalar curvature.

\subsection{General states}

We are going to consider the PDF of the position operator obtained from general states of the harmonic oscillator. Let us consider the basis of the Hamiltonian eigenstates $\left\{|n \rangle\right\} _{n \in \mathcal{N}_0}$, and a state $\hat{\rho}$ of the form
\begin{equation}\label{estado general}
\hat{\rho}= \sum_{n,m}\lambda_{nm}|n \rangle \langle m|.
\end{equation}
The probability distribution function of the position operator is given by
\begin{equation}\label{probabilidad estado general}
P(x)= \langle x|\hat{\rho}|x \rangle = \sum_{n, m}\lambda_{nm}\varphi_n(x)\varphi_m(x)= \sum_{n, m}\frac{\lambda_{nm}a_n a_m}{\sqrt{2\pi}{\sigma}} e^{- \frac{(x-\mu)^2}{2 \sigma^2} } H_n\left(\frac{x-\mu}{\sqrt{2} \sigma}\right) H_m\left(\frac{x-\mu}{\sqrt{2} \sigma}\right),
\end{equation}
where $\varphi_n(x)$ is the wave function of the eigenstate $|n\rangle$, given in equation \eqref{wave function}.

For practical reasons, we define the function $f(y)$,
\begin{equation}\label{funcion f}
f(y)= \sum_{n, m}\frac{\lambda_{nm}a_n a_m}{\sqrt{2\pi}} e^{-y^2} H_n\left(y\right) H_m\left(y\right).
\end{equation}
Then, we have $P(x)= \frac{f(y(x))}{\sigma}$, with $y(x) = \frac{x-\mu}{\sqrt{2} \sigma}$.

In order to calculate the Fisher--Rao metric associated with $P(x)$, we need the partial derivatives $\partial_{\mu}P(x)$ and $\partial_{\sigma}P(x)$, which are given by
\begin{align}
\partial_{\mu}P(x) &= \partial_{\mu} \left(\frac{f(y(x))}{ \sigma}  \right)=\frac{-f'(y(x))}{\sqrt{2} \sigma^2}, \label{P mu} \\
\partial_{\sigma}P(x) &=  \partial_{\sigma}\left( \frac{f(y(x))}{ \sigma} \right)=\frac{-f(y(x))}{ \sigma^2}+ \frac{-y(x) f'(y(x))}{ \sigma^2}, \label{P sigma}
\end{align}
with $f'(y) = \frac{d}{dy}f(y)$.

Replacing the PDF \eqref{probabilidad estado general} and the partial derivatives \eqref{P mu} and \eqref{P sigma} in the integral of equation \eqref{fisherao}, and making the change of variable $y = y(x)$, we obtain the Fisher--Rao metric
\begin{equation}
\begin{split}
I_{\mu \sigma}&=I_{\sigma \mu }= \int_{-\infty}^{+\infty}\frac{\partial_{\mu}P(x)\partial_{\sigma}P(x)}{P(x)}dx =\frac{1}{\sigma^2}\int_{-\infty}^{+\infty}   \left(f'(y) + \frac{ y \left(f'(y)\right)^2 }{f(y)}\right)dy = \frac{1}{\sigma^2}\int_{-\infty}^{+\infty}   \frac{ y \left(f'(y)\right)^2 }{f(y)} dy, \nonumber\\
\nonumber  \\
I_{\mu \mu}&=\int_{-\infty}^{+\infty}\frac{\left(\partial_{\mu}P(x)\right)^2}{P(x)}dx = \frac{1}{\sqrt{2}\sigma^2}\int_{-\infty}^{+\infty} \frac{\left(f'(y)\right)^2}{f(y)}dy, \nonumber \\
\nonumber  \\
I_{\sigma \sigma}&= \int_{-\infty}^{+\infty}\frac{\left(\partial_{\sigma}P(x)\right)^2}{P(x)}dx =\frac{\sqrt{2}}{\sigma^2}\int_{-\infty}^{+\infty}  \frac{\left(f(y)+ y f'(y)\right)^2}{f(y)}dy = \frac{\sqrt{2}}{\sigma^2}\int_{-\infty}^{+\infty}  \left(-f(y)+ 2 (y f(y))' + \frac{y^2 (f'(y))^2}{f(y)}\right)dy = \nonumber\\ & =\frac{\sqrt{2}}{\sigma^2}\int_{-\infty}^{+\infty}  \frac{y^2 (f'(y))^2}{f(y)}dy - \frac{1}{\sigma^2},
\end{split}
\end{equation}
where in the first equation we used that $\int_{-\infty}^{+\infty} f'(y)dy = 0$, and  in the last equation we used that  $\int_{-\infty}^{+\infty}  f(y)dy = \frac{1}{\sqrt{2}}$ and $\int_{-\infty}^{+\infty} (y f(y))'dy = 0$.

Therefore, we can write the Fisher--Rao metric as follows:
\begin{eqnarray}\label{metrica estado general}
I_{\alpha \beta}=\frac{1}{\sigma^2}\left(
\begin{array}{cc}
\tilde{I}_{\mu \mu} & \tilde{I}_{\mu \sigma} \\
\tilde{I}_{\mu \sigma} &  \tilde{I}_{\sigma \sigma} \\
\end{array}
\right),
\end{eqnarray}
where $\tilde{I}_{\mu \sigma}$, $\tilde{I}_{\mu \sigma}$ and $\tilde{I}_{\mu \sigma}$ are independent of $\mu$ and $\sigma$, and they are given by
\begin{equation}
\begin{split}
\tilde{I}_{\mu \sigma}& =\int_{-\infty}^{+\infty}   \frac{ y \left(f'(y)\right)^2 }{f(y)} dy, \nonumber \\
\tilde{I}_{\mu \mu} &=  \frac{1}{\sqrt{2}}\int_{-\infty}^{+\infty} \frac{\left(f'(y)\right)^2}{f(y)}dy, \nonumber \\
\tilde{I}_{\sigma \sigma} &= \sqrt{2}\int_{-\infty}^{+\infty}  \frac{y^2 (f'(y))^2}{f(y)}dy - 1. \nonumber
\end{split}
\end{equation}
From the Fisher--Rao metric and using equations \eqref{christoffel} to \eqref{scalar}, we can obtain the scalar curvature
\begin{eqnarray}\label{curvatura estado general}
R=\frac{2 \tilde{I}_{\mu \mu} }{\tilde{I}_{\mu \sigma}^2- \tilde{I}_{\mu \mu} \tilde{I}_{\sigma \sigma}}.
\end{eqnarray}
The Cram\'er--Rao bound gives the lower covariance matrix of any pair of unbiased estimators $T_1, T_2$ of the parameters $\mu, \sigma$,
\begin{equation}
  \text{cov}\left(T_1, T_2 \right)\geq \frac{\sigma^2}{\tilde{I}_{\mu \mu} \tilde{I}_{\sigma \sigma}-\tilde{I}^2_{\mu \sigma} }  \left(
\begin{array}{cc}
\tilde{I}_{\sigma \sigma}& -\tilde{I}_{\mu \sigma} \\
-\tilde{I}_{\mu \sigma} & \tilde{I}_{\mu \mu} \\
\end{array}
\right).
\end{equation}
Finally, we can express the variance of $T_2$ in terms of the scalar curvature,
\begin{equation}
  \text{var} \left( T_2\right)\geq  -\frac{\sigma^2 R}{2}.
\end{equation}

\noindent \textbf{Corollary 1:}
The Fisher--Rao metric for a general state of the harmonic oscillator is independent of the parameter $\mu$ and it only depends on the parameter $\sigma$ by a general factor $1/\sigma^2$.

~

\noindent \textbf{Corollary 2:}
The scalar curvature for a general state of the harmonic oscillator is independent of the parameters $\mu$ and $\sigma$, and
it only involves integrals of the dimensionless function $f(y)$ and its derivative $f^{\prime}(y)$.

~

\noindent \textbf{Corollary 3:}
The lower variance of unbiased estimators of the parameter $\sigma$ is proportional to $\sigma^2 R$.

\subsection{Mixtures of Hamiltonian eigenstates}

We consider quantum states which are mixtures of the Hamiltonian eigenstates. Mixtures of eigenstates are particular cases of the states given in equation \eqref{estado general}, with $\lambda_{nm}= \delta_{nm}\lambda_{n}$ i.e., $\hat{\rho}= \sum_{n}\lambda_{n}|n \rangle \langle n|$.
Therefore, the probability distribution function of the position operator, the Fisher--Rao metric and the scalar curvature can be obtained from the general expressions \eqref{probabilidad estado general}, \eqref{metrica estado general} and \eqref{curvatura estado general}, considering $\lambda_{nm}= \delta_{nm}\lambda_{n}$.

In this case, the PDF of the position operator takes the form
\begin{equation}
P(x)=  \sum_{n}\lambda_{n}|\varphi_n(x)| =\sum_{n}\lambda_{n}p_n(x|\mu,\sigma).\nonumber
\end{equation}

The diagonal elements of the Fisher--Rao metric are zero, and the elements $I_{\mu \sigma} =I_{\sigma \mu }$ are given in equation \eqref{metrica estado general},
\begin{equation}\label{ap mix g}
I_{\mu \sigma}=I_{\sigma \mu }= \frac{1}{\sigma^2}\int_{-\infty}^{+\infty}   \frac{ y \left(f'(y)\right)^2 }{f(y)} dy.
\end{equation}
with $f(y)= \sum_{n}\frac{\lambda_{n}a^2_n}{\sqrt{2\pi}} e^{-y^2} H^2_n\left(y\right)$.
Since Hermite polynomials $H_n\left(y\right)$ are even or odd functions of the variable $y$, $H^2_n\left(y\right)$ are even functions. Then, $f(y)$ is also an even function and its derivative $f'(y)$ is an odd function. Finally, the integrand of equation \eqref{ap mix g} is an odd function of $y$. Therefore, $I_{\mu \sigma}=I_{\sigma \mu } = 0$,

Finally, the scalar curvature is obtained from equation \eqref{curvatura estado general},
\begin{eqnarray}
R=- \frac{2}{\tilde{I}_{\sigma \sigma}}. \nonumber
\end{eqnarray}

As an example, we consider the mixture state $\hat{\rho}_{01}= \frac{1}{2}|0 \rangle \langle 0| + \frac{1}{2}|1 \rangle \langle 1|$.
The Fisher--Rao metric is given by
\begin{eqnarray}
I^{(01)}_{\alpha \beta}=\frac{1}{\sigma^2}\left(
\begin{array}{cc}
2 + \sqrt{2 e \pi} \left(\text{Erf}\left(\frac{1}{\sqrt{2}}\right)-1\right) & 0 \\
0 &  2 + \sqrt{2 e \pi} \left(1- \text{Erf}\left(\frac{1}{\sqrt{2}}\right)\right)  \\
\end{array}
\right),
\end{eqnarray}
where $\text{Erf}(x)$ is the Gauss error function, with $\text{Erf}\left(\frac{1}{\sqrt{2}}\right) \approx 0.317$.
The scalar curvature is approximately $R^{(01)} \approx-0.604$.

\subsection{Superposition of Hamiltonian eigenstates}

We consider quantum states which are superpositions of Hamiltonian eigenstates.
Superpositions of eigenstates of the form $|\psi \rangle = \sum_{n} \alpha_n |n\rangle$ are particular cases of states given in equation \eqref{estado general}, with $\lambda_{nm}= \alpha_n \alpha^*_m$, i.e., $\hat{\rho}= |\psi \rangle \langle \psi|= \sum_{nm}\alpha_n \alpha^*_m |n \rangle \langle m|$.
Therefore, the PDF of the position operator, the Fisher--Rao metric and the scalar curvature can be obtained from the general expressions \eqref{probabilidad estado general}, \eqref{metrica estado general} and \eqref{curvatura estado general}, considering $\lambda_{nm}= \alpha_n \alpha^*_m$

\subsubsection{Even or odd superpositions}
In this section we focus on a family of superpositions that yield
analytic expressions. If we consider a superposition of eigenstates
with only even or odd eigenstates, i.e.,

\begin{equation}
\hat{\rho}= \sum_{\substack{n, m \\ \text{ even indexes}}} \alpha_n \alpha^*_m |n \rangle \langle m|,~~~ \text{ or } ~~~\hat{\rho}= \sum_{\substack{n, m \\ \text{ odd indexes}}} \alpha_n \alpha^*_m |n \rangle \langle m|, \nonumber
\end{equation}

\noindent we obtain that the diagonal elements of the Fisher--Rao
metric are zero. The proof is similar to the case of mixtures of
eigenstates. The diagonal elements are given in equation
\eqref{metrica estado general},
\begin{equation} \label{superposicion g mu sigma}
I_{\mu \sigma}=I_{\sigma \mu }= \frac{1}{\sigma^2}\int_{-\infty}^{+\infty}   \frac{ y \left(f'(y)\right)^2 }{f(y)} dy,
\end{equation}
with
\[
f(y)= \sum_{\substack{\text{even or odd} \\ \text{indexes}}}\frac{\alpha_n \alpha^*_m a_n a_m}{\sqrt{2\pi}} e^{-y^2} H_n\left(y\right) H_m\left(y\right).
\]
If the indexes $n, m$ can only take even or odd values, then the product $H_n\left(y\right) H_m\left(y\right)$ is always an even function of the variable $y$. Then, $f(y)$ is also an even function and its derivative $f'(y)$ is an odd function. Finally, the integrand of equation \eqref{superposicion g mu sigma} is an odd function of $y$, and the result of the integral is zero.

Again, we obtain that the scalar curvature, given in equation \eqref{curvatura estado general}, is
\begin{eqnarray}
R=- \frac{2}{\tilde{I}_{\sigma \sigma}}. \nonumber
\end{eqnarray}

\subsubsection{Real or imaginary superpositions}

Analytic expressions can also be obtained for superpositions of
eigenstates that involve only real coefficients, i.e., $\hat{\rho}=
\sum_{n, m } \alpha_n \alpha_m |n \rangle \langle m|$. In order to
compute the Fisher--Rao metric, we need the fuction $f(y)$, given in
\eqref{funcion f}, and it derivative $f'(y)$,

\begin{equation}\label{funcion f caso 2}
\begin{split}
f(y)&=  \frac{e^{-y^2}}{\sqrt{2\pi}}\left(\sum_{n}\alpha_n a_n  H_n\left(y\right)  \right)^2, \\
f'(y)&= \frac{2 e^{-y^2}}{\sqrt{2\pi}} \left(\sum_{n}\alpha_n a_n  H_n\left(y\right) \right) \left[ \sum_{n}\alpha_n a_n  \left(n H_{n-1}(y)- \frac{H_{n+1}(y)}{2} \right)\right],
\end{split}
\end{equation}

\noindent where in the last equation we have used the recurrence
relations of the Hermite polynomials \eqref{recurrencia}. Replacing
expressions \eqref{funcion f caso 2} in the Fisher--Rao metric
\eqref{metrica estado general}, and taking into account relations
\eqref{orthogonality} and \eqref{recurrencia}, we obtain
\begin{equation}
\begin{split}
I_{\mu \sigma}= I_{\sigma \mu } &= \frac{1}{\sigma^2}\int_{-\infty}^{+\infty}   \frac{4 y e^{-y^2}}{\sqrt{2\pi}} \left[ \sum_{n}\alpha_n a_n  \left(n H_{n-1}(y)- \frac{H_{n+1}(y)}{2} \right)\right]^2 dy=  \nonumber \\
& =\frac{1}{\sigma^2} \sum_{n}\alpha_n \left( \alpha_{n-3} \sqrt{n (n-1)(n-2)}+ \alpha_{n-1} n\sqrt{n} + \alpha_{n+1} (n+1)\sqrt{n +1} + \alpha_{n+3} \sqrt{(n+3)(n+2) (n+1)} \right),  \nonumber \\
\nonumber \\
I_{\mu \mu} &=  \frac{1}{\sigma^2}\int_{-\infty}^{+\infty} \frac{2e^{-y^2}}{\sqrt{\pi}} \left[ \sum_{n}\alpha_n a_n  \left(n H_{n-1}(y)- \frac{H_{n+1}(y)}{2} \right)\right]^2 dy=  \nonumber \\
&= \frac{1}{\sigma^2} \sum_{n}\alpha_n\left(-  \alpha_{n-2}   \sqrt{ n  (n-1)}+ \alpha_n (2n+1) - \alpha_{n+2} \sqrt{(n+2)(n+1)} \right),  \nonumber\\
\nonumber \\
I_{\sigma \sigma} &=  \frac{1}{\sigma^2}\int_{-\infty}^{+\infty}   \frac{4 y^2 e^{-y^2}}{\sqrt{\pi}} \left[ \sum_{n}\alpha_n a_n  \left(n H_{n-1}(y)- \frac{H_{n+1}(y)}{2} \right)\right]^2 dy -  \frac{1}{\sigma^2} =  \nonumber \\
& =\frac{1}{\sigma^2} \sum_{n}\alpha_n \left(-\alpha_{n-4}\sqrt{n (n-1)(n-2)(n-3)} + \alpha_n ( 2n^2 +2n +3)-\alpha_{n+4}\sqrt{(n+4)(n+3)(n+2)(n+1)} \right) -  \frac{1}{\sigma^2}.  \nonumber
\end{split}
\end{equation}

If we consider a superposition of eigenstates with only imaginary
coefficients, we obtain a similar result, but replacing the
coefficients $\alpha_n$ by its imaginary part, i.e., by
Im$(\alpha_n)$.


\section{Conclusions}

In this work we have proposed a generalization of the Gaussian model
-namely, the \emph{Hermite-Gaussian model}- and we have studied many of its
properties from the point of view of the information geometry
approach. We have shown its relation with the probabilities associated to
the one-dimensional quantum harmonic oscillator model and
analytic expressions for some particular classes of states were provided.
Specifically, we found that for finite mixtures of eigenstates
and finite superpositions
of (even or odd) eigenstates the Fisher metric is
always diagonal.
Real and imaginary superpositions of eigenstates
do not imply a diagonal Fisher metric and the matrix elements
are given in terms of a series sum.
An analytic expression for the scalar curvature was only obtained when
the Fisher metric is diagonal, being negative and inversely proportional to the
$\sigma\sigma$ element.

Due to
the relevance of this model in many applications, our contribution
may serve to extend the scope of information geometry techniques
into a wider class of physical problems.
For example, since in irreversible
processes the final (reduced) state of a system (after interacting
with the environment)
is typically a mixture of their eigenstates, which
for the case of Hermite-Gaussian models
has a diagonal Fisher metric,
the results obtained could be used for
determining if
the process involved is irreversible or not
by simple inspection of the diagonal elements of the Fisher
metric. In this context and considering that
the states of the system
can be expressed by means of Hermite-Gaussian models,
if the Fisher metric of the final reduced
state results non-diagonal then by Sections 3.3 and 3.4 it is not mixture of
harmonic oscillator eigenstates,
and thus the process cannot be irreversible.


\section*{ACKNOWLEDGMENTS}
This research was founded by the CONICET,
CAPES / INCT-SC (at
Universidade Federal da Bahia, Brazil),
the National University of
La Plata and the University of Buenos Aires.

\appendix
\section{Hermite polynomials}\label{apendice1}
The Hermite polynomials $H_n$ are given by the expression
\begin{equation}
H_n (y) = (-1)^n e^{y^2} \frac{d^n }{dy^n} e^{-y^2}, \nonumber
\end{equation}
and their orthogonality relation is
\begin{equation}
\label{orthogonality}
\int_{-\infty}^{+\infty} e^{-y^2} H_n (y) H_m (y)dy= \sqrt{\pi}\,  2^n\, n! \, \delta_{n,m}.
\end{equation}
An important feature of these polynomials is that if $n$ is even, $H_n(y)$ is an even function; and if $n$ is odd, $H_n(y)$ is an odd function.

Some relevant recurrence relations are the following:
\begin{equation}
H'_{n}(y) = 2 n H_{n-1}(y),  ~~~~~~ H_{n+1}(y) = 2 y H_n(y) -2 n H_{n-1}.  \label{recurrencia}
\end{equation}

\section{Hermite--Gaussian model}\label{apendice2}

For parameters $\mu$ and $\sigma$, the probability distribution of the $n$--Hermite--Gaussian model is
\begin{equation}
\label{probabilidad de y}
p_n (x)= \frac{1}{\sqrt{2\pi}\sigma}e^{-y^2}a^2_n H^2_n\left(y\right), ~~~~~\text{with} ~~~~~ a_n = \frac{1}{\sqrt{2^n n!}}, ~~~~~ y =\frac{x -\mu}{\sqrt{2}\sigma}.
\end{equation}
In order to obtain the elements of the metric tensor, we need to calculate the partial derivatives of the probability distribution. It easy to show that
\begin{align}
\partial_\mu p_{n}(x) &=-\frac{p_n^{\prime}(y)}{\sqrt{2} \,\sigma}, \label{partial1} \\
\partial_\sigma p_{n}(x) &=-\frac{p_n(y) + y p_n^{\prime}(y)}{\sigma},  \label{partial2}
\end{align}
with
\begin{equation}
p_n^{\prime}(y)=\frac{dp_n}{dy}(y) = \frac{2 a_n^2}{\sqrt{2\pi}\sigma} \,  e^{-y^2}H_{n}(y) \left(n H_{n-1}(y) -  \frac{1}{2} H_{n+1}(y) \right),  \label{derivada de p}
\end{equation}
where we have used the recurrence relations (\ref{recurrencia}). It should be noted that $p_n(y)$ is even an function of $y$, thus $p_n^{\prime}(y)$ is an odd function of $y$.

Also, we will need to express $y p_n^{\prime}(y)$ in terms of Hermite polynomials,
\begin{equation}
\begin{split}
\label{y derivada de p}
y p_n^{\prime}(y)  &=  \frac{2 a_n^2}{\sqrt{2\pi}\sigma} \,  e^{-y^2}H_{n}(y) \left(n y H_{n-1}(y) -  \frac{1}{2} y H_{n+1}(y) \right)= \nonumber \\
&=  \frac{2 a_n^2}{\sqrt{2\pi}\sigma} \,  e^{-y^2}H_{n}(y) \left(n (n-1)  H_{n-2}(y) - \frac{1}{2} H_n(y) - \frac{1}{4} H_{n+2}(y) \right),
\end{split}
\end{equation}
where we have used expression (\ref{derivada de p}) and the recurrence relations (\ref{recurrencia}).

\subsection{Off--diagonal elements}

Since the metric tensor is symmetric, it is enough to calculate the element $I^{(n)}_{\mu \sigma}$, given by
\begin{equation}
\label{metrica mu sigma}
I^{(n)}_{\mu \sigma} = \int_{-\infty}^{+\infty}\frac{1}{p_{n}(x)} \partial_\mu p_{n}(x) \partial_\sigma p_{n}(x) dx.
\end{equation}
Replacing expressions (\ref{partial1}) and (\ref{partial2}) in (\ref{metrica mu sigma}) and doing some easy manipulations, we obtain
\begin{equation}
I^{(n)}_{\mu \sigma} = \int_{-\infty}^{+\infty}\frac{1}{\sigma^2} \left(p_n^{\prime}(y(x)) + y(x) \frac{\left[p'_{n}(y(x))\right]^2}{p_n(y(x))} \right) dx =  \int_{-\infty}^{+\infty}\frac{\sqrt{2}}{\sigma} \left(p_n^{\prime}(y) + y \frac{\left[p'_{n}(y)\right]^2}{p_n(y)} \right) dy, \label{integrando}
\end{equation}
where in the last equation we changed from variable $x$ to the variable $y =\frac{x -\mu}{\sqrt{2}\sigma}$.
Since $p_n(y)$ and $p_n^{\prime}(y)$ are even and odd functions of $y$, respectively, then the integrand of (\ref{integrando}) is an odd function. Therefore, $I^{(n)}_{\mu \sigma} = 0$.

\subsection{Element $I^{(n)}_{\mu \mu}$}

The element $I^{(n)}_{\mu \mu}$ is given by

\begin{equation}
\label{metrica mumu}
I^{(n)}_{\mu \mu} = \int_{-\infty}^{+\infty}\frac{1}{p_{n}(x)} \left[\partial_\mu p_{n}(x)\right]^2  dx.
\end{equation}
Replacing expression (\ref{partial1}) in (\ref{metrica mumu}), we obtain
\begin{equation}
\label{metrica mumu2}
I^{(n)}_{\mu \mu} = \int_{-\infty}^{+\infty}\frac{1}{2 \,\sigma^2}\frac{\left[ p_n^{\prime}(y(x))\right]^2}{p_n(y(x))}  dx =\int_{-\infty}^{+\infty}\frac{1}{ \sqrt{2}\sigma}\frac{  \left[p_n^{\prime}(y)\right]^2} {p_n(y)} dy.
\end{equation}
In the last step, we have changed from variable $x$ to the variable $y$. Then, if we replace expressions (\ref{probabilidad de y}) and (\ref{derivada de p}) in (\ref{metrica mumu2}) and we rearrange the expression, we obtain
\begin{equation}
\begin{split}
I^{(n)}_{\mu \mu} &=\frac{2 a_n^2}{\sqrt{\pi} \sigma^2} \left[ n^2 \int_{-\infty}^{+\infty} e^{-y^2}  H^2_{n-1}(y) dy - n\int_{-\infty}^{+\infty} e^{-y^2} H_{n-1}(y) H_{n+1}(y)  dy  + \frac{1}{4}\int_{-\infty}^{+\infty} e^{-y^2}H^2_{n+1}(y)  dy \right] = \nonumber \\
&=\frac{2 }{\sqrt{\pi} \sigma^2} \frac{1}{2^n \, n!} \left( n^2 \sqrt{\pi}\,  2^{n-1}\, (n-1)!  + \frac{1}{4}  \sqrt{\pi}\,  2^{n+1}\, (n+1)! \right).  \nonumber
\end{split}
\end{equation}
In the last step we have used the orthogonality relation (\ref{orthogonality}). Finally, we obtain $I^{(n)}_{\mu \mu}= \frac {2n +1} {\sigma ^2}$.

\subsection{Element $I^{(n)}_{\sigma \sigma}$}
The element $I^{(n)}_{\mu \mu}$ is given by
\begin{equation}
\label{metrica sigmasigma}
I^{(n)}_{\sigma \sigma} = \int_{-\infty}^{+\infty}\frac{1}{p_{n}(x)} \left[\partial_\sigma p_{n}(x)\right]^2  dx.
\end{equation}
Replacing expression (\ref{partial2}) in (\ref{metrica sigmasigma}), we obtain
\begin{equation}
\label{metrica sigmasigma2}
I^{(n)}_{\sigma \sigma} =\int_{-\infty}^{+\infty}\frac{1}{ p_n(y(x))} \left( -\frac{p_n(y(x)) + y(x) p_n^{\prime}(y(x))}{\sigma}\right)^2 dx =\int_{-\infty}^{+\infty}\frac{\sqrt{2}}{ \sigma}\frac{  \left[ p_n(y)+y p_n^{\prime}(y)\right]^2} {p_n(y)} dy.
\end{equation}
In the last equation we have changed from variable $x$ to the variable $y$. Then, if we replace expressions (\ref{probabilidad de y}) and (\ref{y derivada de p}) in (\ref{metrica sigmasigma2}) and we rearrange the expression, we obtain
\begin{align}
I^{(n)}_{\sigma \sigma} &=\int_{-\infty}^{+\infty}  \frac{a_n^2}{\sqrt{ \pi} \sigma^2} \,  e^{-y^2}\left(2 n(n-1) H_{n-2}(y) - \frac{1}{2}H_{n+2}(y) \right)^2  dy= \nonumber \\
&=  \int_{-\infty}^{+\infty} \frac{a_n^2}{\sqrt{ \pi} \sigma^2} \, e^{-y^2} \left( 4 n^2 (n-1)^2   H_{n-2}^2(y)  + \frac{1}{4} H_{n+2}^2(y) - 2n(n-1)  H_{n-2}(y)H_{n+2}(y) \right)dy=  \nonumber \\
&= \frac{1}{\sqrt{ \pi} \sigma^2}\frac{1}{2^n \, n!} \left( 4 n^2 (n-1)^2 \sqrt{\pi} \, 2^{n-2} \, (n-2)! + \frac{1} {4}\sqrt{\pi} \, 2^{n+2} \, (n+2)! \right). \nonumber
\end{align}
In the last step, we have used the orthogonality relation
(\ref{orthogonality}). Finally, we obtain $I^{(n)}_{\sigma \sigma}
=\frac{2(n^2 +n+1)}{\sigma ^2}$.


\section*{References}

\begin{thebibliography}{99}

\bibitem{fisher}
R. Fisher,
\emph{Phil.
Trans. R. Soc. Lond. A} \textbf{222}, 309-368 (1922).

\bibitem{frieden} R. Frieden. \emph{Physics from Fisher information: a unification}; Cambridge,
UK, Cambridge University Press, 1998.

\bibitem{jaynes} Jaynes, E. T. Information Theory and Statistical Mechanics, I, II.
{\em Phys. Rev.} {\bf 1957}, {\em 106}, 620--630; {\em 108}, 171--190.




\bibitem{rao} C. R. Rao, Bull. Calcutta Math. Soc. \textbf{37}, 81 (1945).

\bibitem{rao2} C. R. Rao, {\em Differential Geometry in Statistical Inference}. In chap. Differential metrics in probability spaces; Institute of Mathematical Statistics, Hayward, CA, 1987.

\bibitem{weinhold} F. Weinhold, \emph{J. Chem. Phys.} \textbf{63}, 2479, 2488 (1975); \textbf{65}, 559 (1976).

\bibitem{ruppeiner} \emph{Rev. Mod. Phys.} \textbf{67}, 605 (1995).

\bibitem{amari} S. Amari, H. Nagaoka. {\em Methods of Information Geometry}; Oxford University Press: Oxford, UK, 2000.

\bibitem{ingarden} R. S. Ingarden, Tensor N.S. 30, 201 (1976).

\bibitem{janyszek} H. Janyszek, \emph{Rep. Math. Phys.} \textbf{24}, 1, 11 (1986).

\bibitem{gibbs} J. W. Gibbs. \emph{The collected works}, \textbf{Vol. 1}, Thermodynamics, Yale University
Press, 1948.

\bibitem{hermann} R. Hermann. \emph{Geometry, physics and systems}, Dekker, New York, 1973.

\bibitem{mrugala} R. Mruga{\l}a.
\emph{Rep. Math. Phys.} \textbf{14}, 419 (1978).

\bibitem{ruppeiner2}  G. Ruppeiner,
\emph{Phys. Rev. A} \textbf{63}, 20 (1979).

\bibitem{carateodory} C. Carathe\'{o}dory.
\emph{Untersuchungen \"{u}ber die Grundlagen der Thermodynamik},
Gesammelte Mathematische Werke, Band 2, Munich, 1995.

\bibitem{bures} D. Bures, \emph{Trans. Am. Math. Soc.} \textbf{135}, 199 (1969).

\bibitem{geert} G. Verdoolaege, \emph{AIP Conf. Proc.} \textbf{1641}, 564--571 (2014).

\bibitem{geert2} G. Verdoolaege, \emph{Rev. Sci. Instrum.} \textbf{85}, 11E810 (2014).

\bibitem{tsallis} C. Tsallis, \emph{J. Stat. Phys.} \textbf{52}, 479 (1988).

\bibitem{abe} S. Abe, \emph{Phys. Rev. E} \textbf{68}, 031101 (2003).

\bibitem{naudts} J. Naudts, \emph{Open Sys. and Information Dyn.} \textbf{12}, 13 (2005).

\bibitem{portesi} M. Portesi, A. Plastino, F. Pennini, \emph{Physica A} \textbf{365}, 173--176 (2006).

\bibitem{portesi2} M. Portesi, A. Plastino, F. Pennini,  \emph{Physica A} \textbf{373}, 273--282 (2007).


\bibitem{cafaro} C. Cafaro, \emph{Chaos Solitons \& Fractals} \textbf{41}, 886--891 (2009).
\bibitem{cafaro2} C. Cafaro, S. Mancini, \emph{Physica D}  \textbf{240}, 607--618 (2011).

\bibitem{cafaro3} C. Cafaro, A. Giffin, C. Lupo, S. Mancini, \emph{Open Syst. Inf. Dyn.} \textbf{19}, 1250001 (2012).

\bibitem{cafaro4}
D.-H. Kim, S. A. Ali, C. Cafaro, S. Mancini,
\emph{Physica A} \textbf{391}, 4517--4556
(2012).

\bibitem{cafaro5} A. Giffin, S. A. Ali, C. Cafaro, \emph{Entropy} \textbf{15}, 4622-4623 (2013).


\bibitem{nacho1} I. S. Gomez, \emph{Physica A} \textbf{484}, 117--131 (2017).

\bibitem{nacho2} I. S. Gomez, M. Portesi,
\emph{AIP Conf. Proc.} \textbf{1853}, 100001 (2017).

\end{thebibliography}
\end{document}